\documentclass[proof]{WileyASNA-v1}

\articletype{Article Type}%

\received{17 January 2023}
\revised{.. ...  2023}
\accepted{.. ...  2023}

\begin{document}

\title{Optical spectroscopy of Be stars:  
 peak separation of Balmer emission lines}

\author[1]{Radoslav K. Zamanov*}
\author[1]{Kiril A. Stoyanov}
\author[1]{Stefan Y. Stefanov}
\author[2,3]{Michael F. Bode} 
\author[1]{Milen S. Minev}


\authormark{Zamanov, Stoyanov, Stefanov, Bode, Minev}

\address[1]{\orgdiv{Institute of Astronomy and National Astronomical Observatory}, \orgname{Bulgarian Academy of Sciences}, \orgaddress{\state{72 Tsarigradsko Shose, 1784 Sofia}, \country{Bulgaria}}}

\address[2]{\orgdiv{Astrophysics Research Institute}, \orgname{Liverpool John Moores University}, 
\orgaddress{\state{IC2, 149 Brownlow Hill, Liverpool, L3 5RF}, \country{UK}}}

\address[3]{\orgdiv{Office of the Vice Chancellor}, \orgname{Botswana International University of Science and Technology}, \orgaddress{\state{Palapye}, \country{Botswana}}}

\corres{* \email{rkz@astro.bas.bg}}


\abstract{The Be stars display variable optical emission lines originating in the circumstellar disc. Here we analyse high resolution spectroscopic observations of Be stars  
and the distance between the peaks of  
 H$\alpha$,  H$\beta$, and  H$\gamma$  emission lines   ($\Delta V_\alpha$, 
  $\Delta V_\beta$,  and $\Delta V_\gamma$ respectively). 
 Combining published data, spectra from the ELODIE archive (obtained in the period 1998 -- 2003)
   and Rozhen spectra (obtained 2015 -- 2023) of 93 Be stars, 
  we find a set of relations  connecting 
  $\Delta V_\alpha$, $\Delta V_\beta$ and $\Delta V_\gamma$. 
  They are effective for  
  $30 \le \Delta V_\alpha \le 500$~km~s$^{-1}$,
  $80 \le \Delta V_\beta  \le 600$~km~s$^{-1}$,  
  and 
  $40 \le \Delta V_\gamma \le 300$~km~s$^{-1}$.
  The new equations are in the form $y=ax + b$ and are valid for 
  a wider velocity range than in previous studies. 
}


\fundingInfo{Bulgarian National Science Fund -- project K$\Pi$-06-H28/2 08.12.2018  "Binary stars with compact object"}

\maketitle


\section{Introduction}
The Be stars are fast-rotating B-type stars
which, at some point in their lives, have shown spectral lines
in emission \citep{2003PASP..115.1153P}. 
They are main sequence or evolved stars, 
belonging to luminosity classes III -- V and having
masses and radii ranging between M $\sim$ 3.6 -- 20~M$_\odot$ and R $\sim$ 2.7 -- 15~R$_\odot$ \citep{2000PhT....53j..77C}.
In the optical spectra, the most prominent observational characteristics
of Be stars are  variable emission lines of different chemical elements such 
as hydrogen, helium, iron, etc. The emission lines
may change their intensity and even disappear during the star's life.
In addition, an infrared excess is present in their continuum \citep{1974ApJ...191..675G}. 
The emission lines and the infrared excess indicate the presence of a geometrically thin, equatorial, 
gaseous, decretion disc which orbits the star in near-Keplerian rotation 
\citep{2013A&ARv..21...69R}. 
Interferometric observations confirm 
the disc-like structure of the circumstellar material 
\citep{2007A&A...464...59M, 2019A&A...621A.123C}).
The emission lines can be symmetric or asymmetric, with double-peak or more complicated 
 structure  \citep{1995A&A...302..751H,2010ApJS..187..228S} 
and the peaks' separation can be used for an estimation of the disc radius 
and to test theoretical models.


%
%

 \begin{figure*}    
  \vspace{12.0cm}   
  \includegraphics{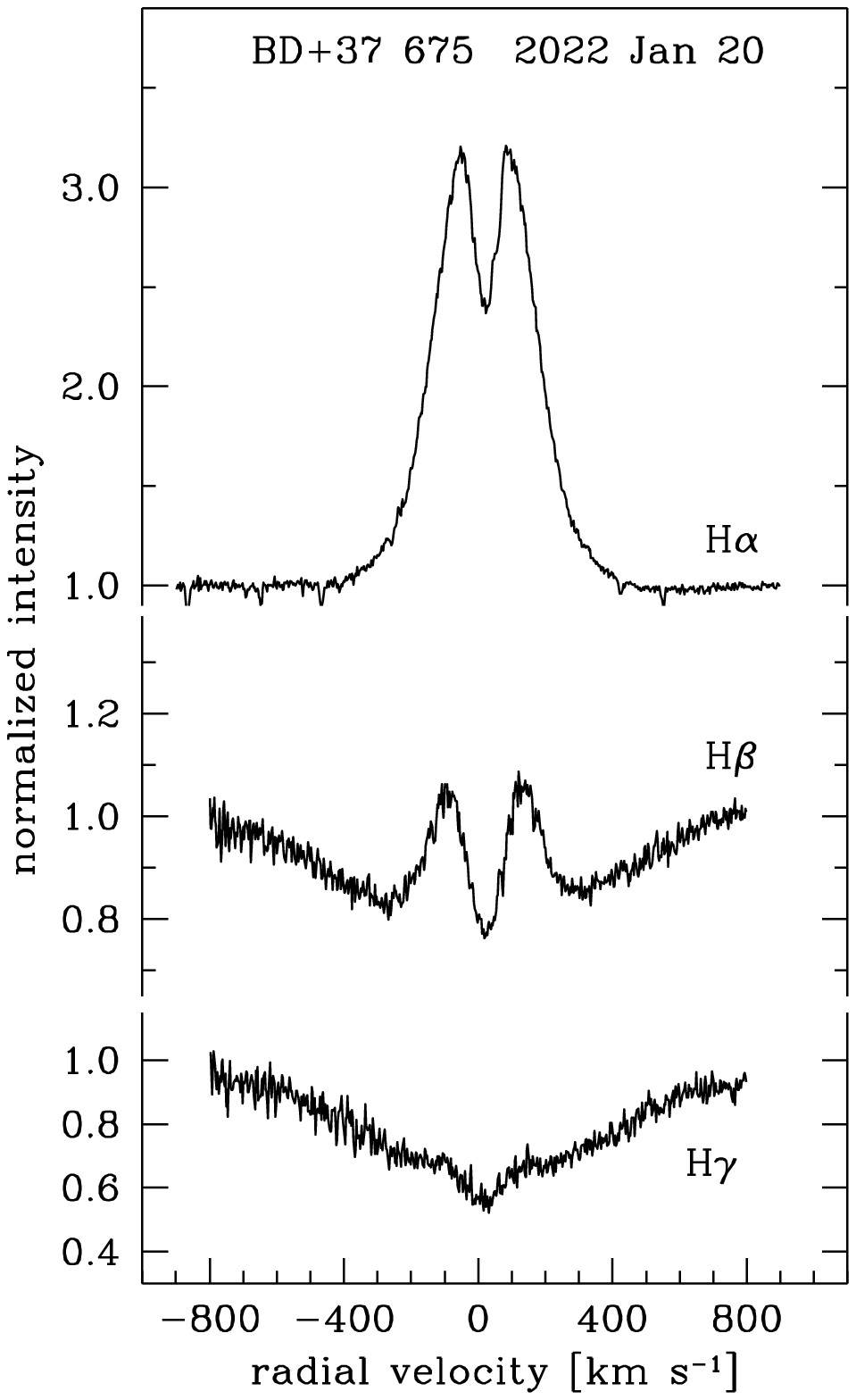}	  
  \includegraphics{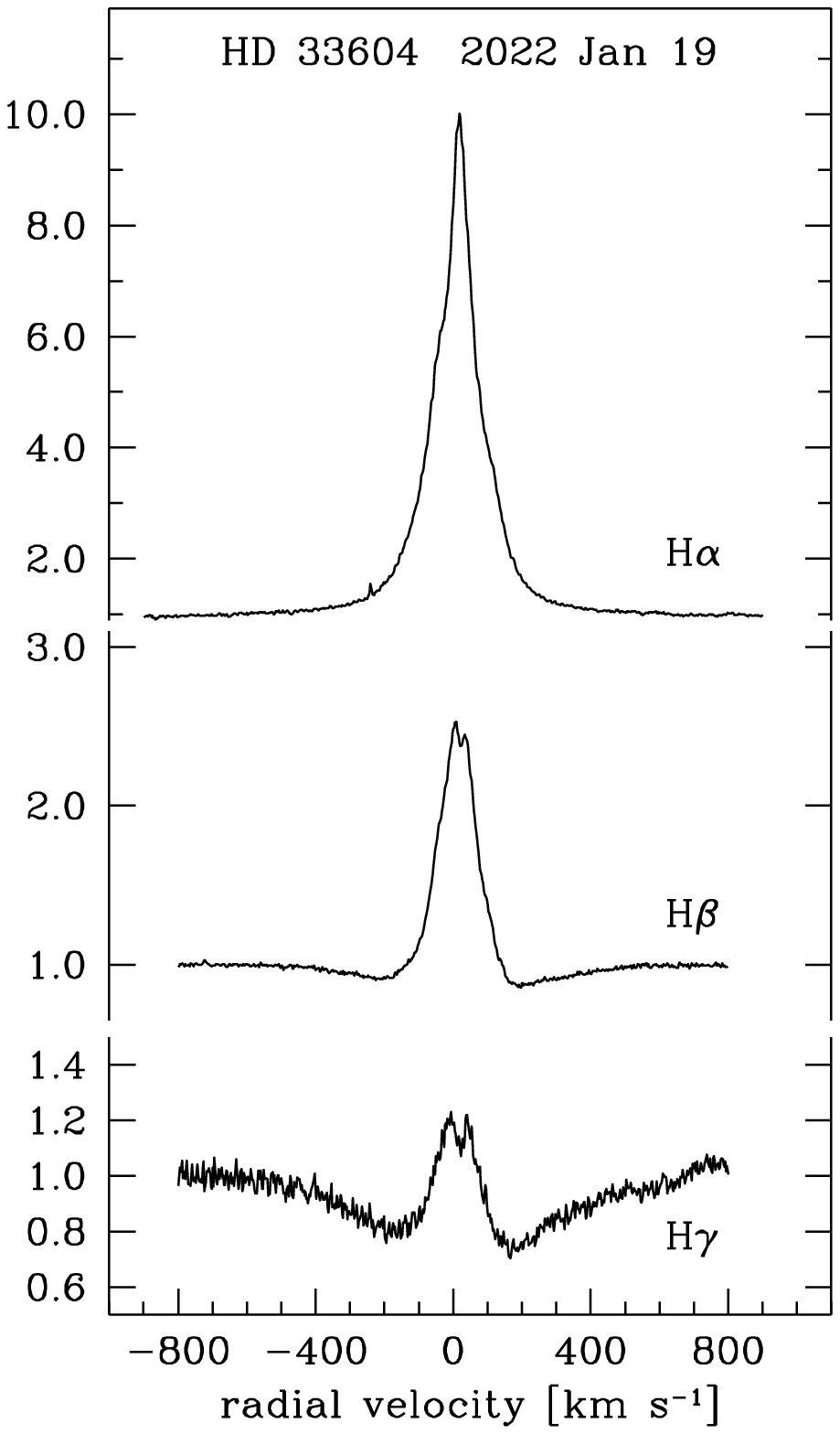}
  \includegraphics{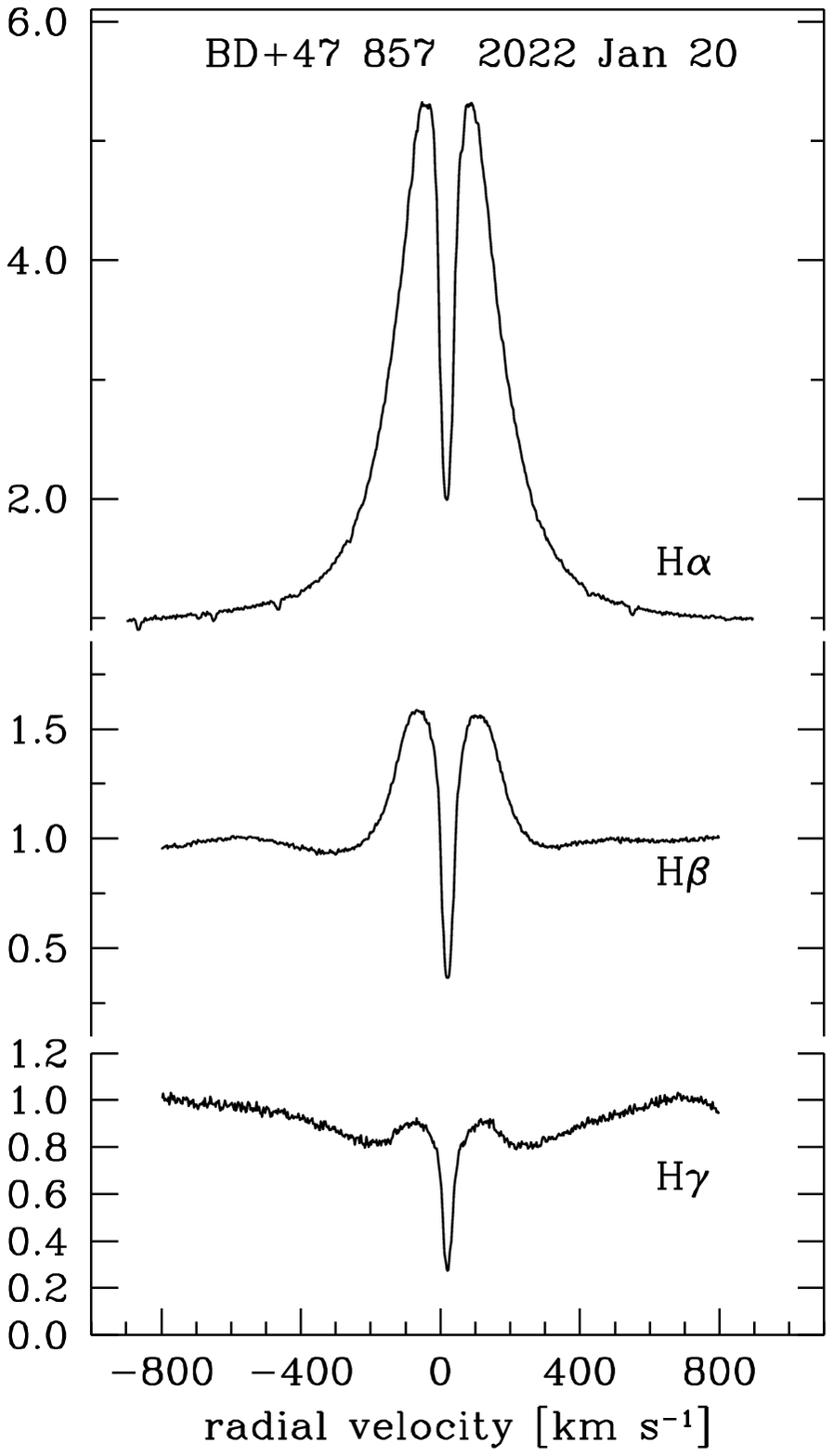}	  
  \caption[]{Examples of the profiles of H$\alpha$, H$\beta$, and H$\gamma$ lines.
	     The object and the date of observation are marked at the top of each panel. }
\label{f.ex}      
\end{figure*}	     


  \begin{figure}[ht!] 
  \vspace{8.9cm} 
  \includegraphics{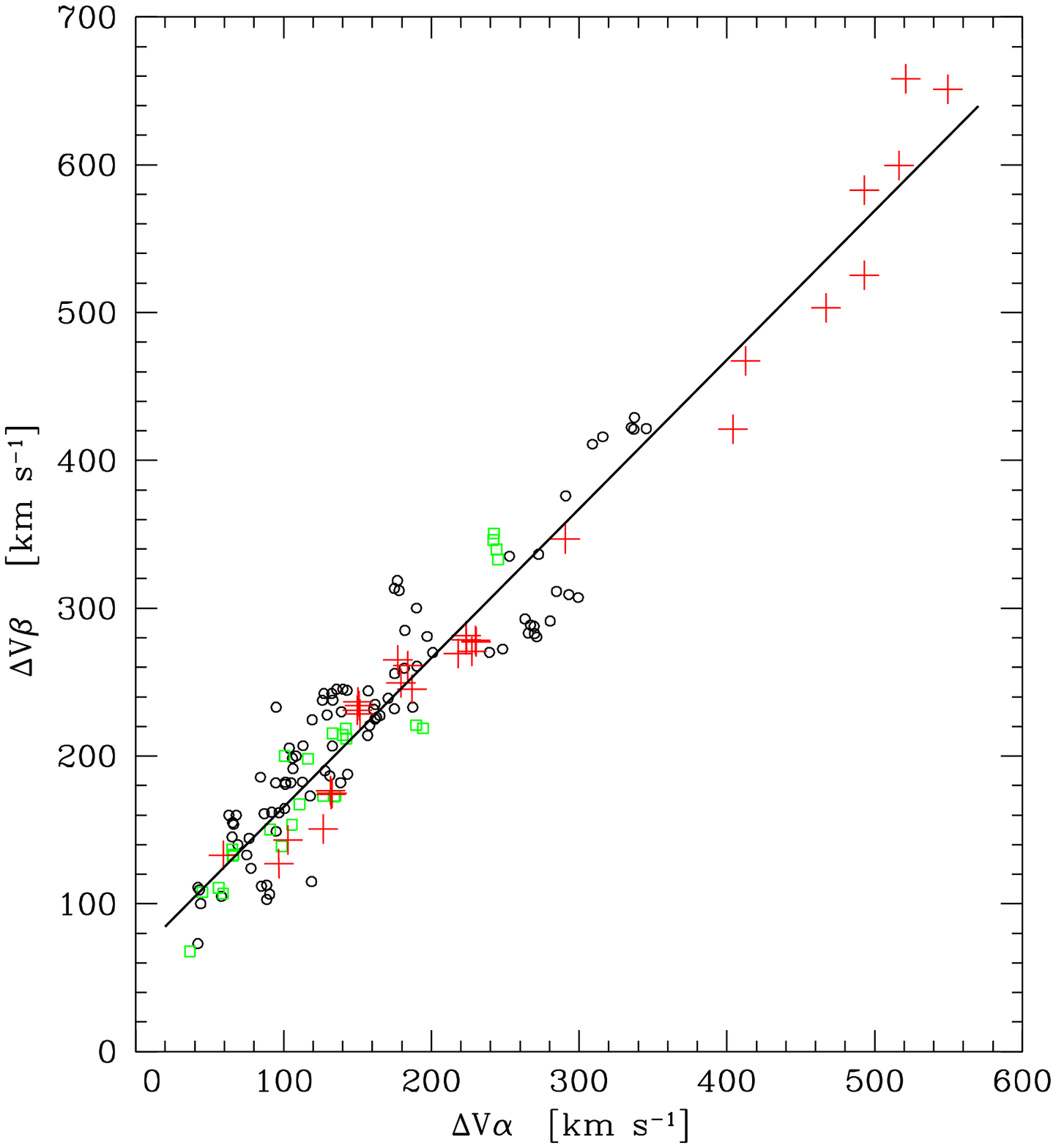} 	  
  \caption[]{ $\Delta V_\beta$ versus  $\Delta V_\alpha$ for Be stars.
  The black circles are data from the literature, 
  the green squares are ELODIE data, the red pluses are Rozhen data.  
  The solid line is the linear fit 
   $\Delta V_\beta = 1.01 \; \Delta V_\alpha + 64.2$ km~s$^{-1}$.}
   \label{f.ab}  
  \vspace{8.9cm}  
  \includegraphics{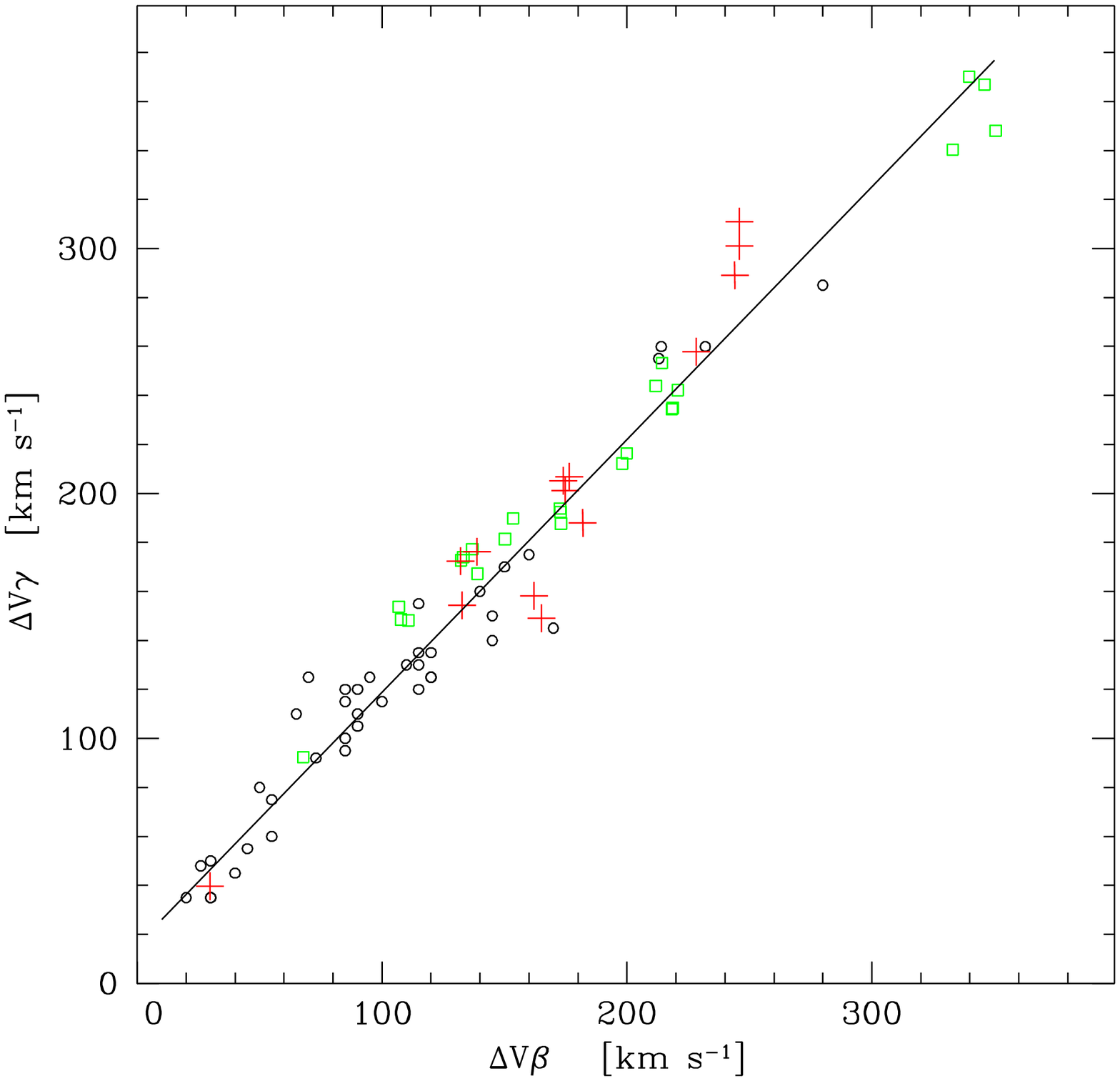} 	  
  \caption[]{ $\Delta V_\gamma$ versus  $\Delta V_\beta$ for Be stars.
     The symbols are as in Fig.~\ref{f.ab}. 
     The black solid line is the linear fit 
     $\Delta V_\gamma = 1.03 \; \Delta V_\beta + 15.7$ km~s$^{-1}$.  
     } 
  \label{f.bg}  
\end{figure}

  \begin{figure}[ht!]  
   \vspace{8.9cm} 
  \includegraphics{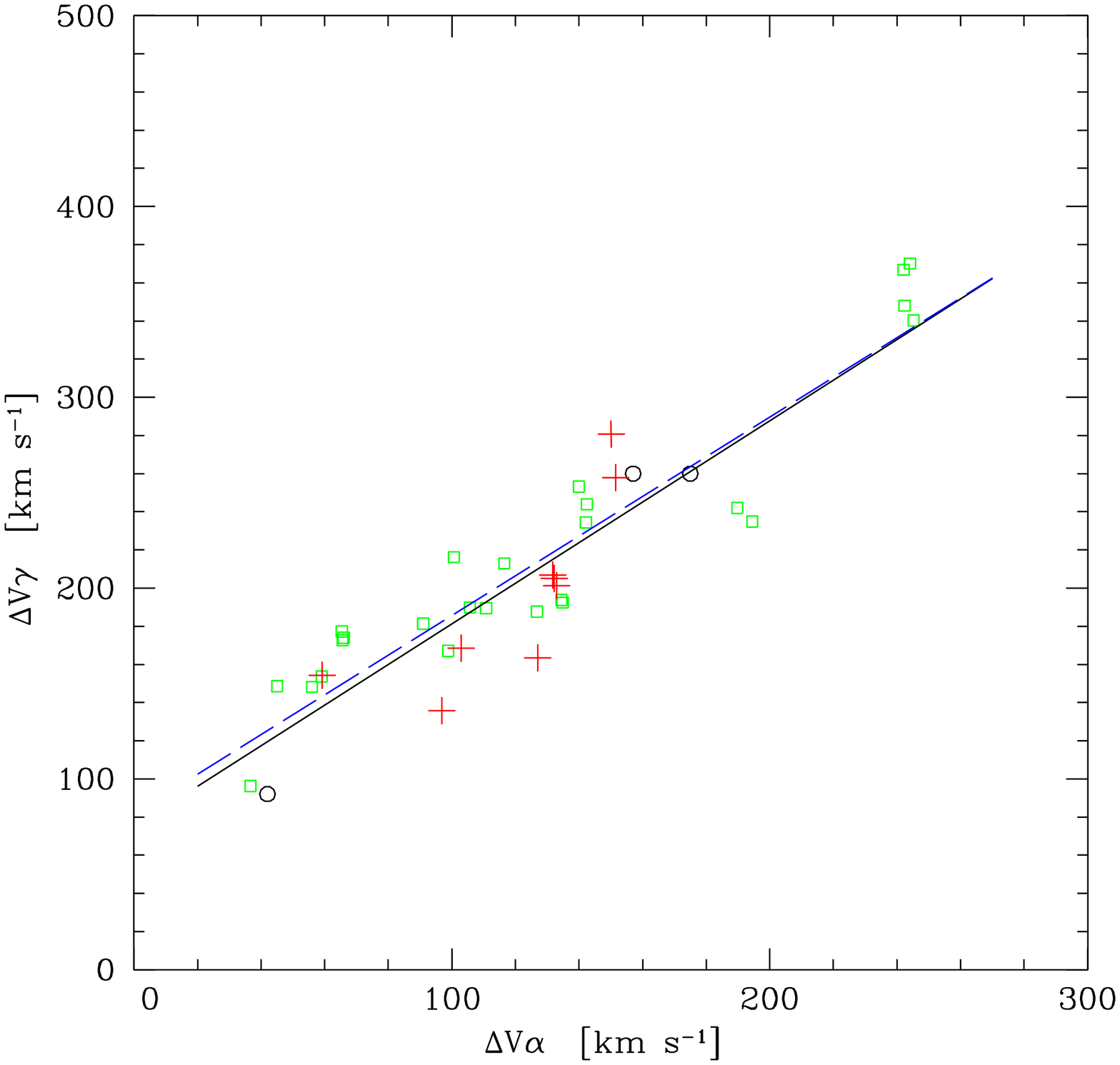} 	  
  \caption[]{ $\Delta V_\gamma$ versus  $\Delta V_\beta$ for Be stars.
     The symbols are the same as in Fig.~\ref{f.ab}, and
     represent the spectra on which 
     both $\Delta V_\gamma$ and $\Delta V_\alpha$
     are measured simultaneously.  
     The blue dashed line is Eq.~\ref{eq.ag1}, 
     and the black solid line is the fit as given by Eq.~\ref{eq.ag2}. 
     } 
  \label{f.ag}  
\end{figure}

 %
 %

In this note, we analyse the separation of the peaks 
of the Balmer emission lines formed in the circumstellar disc 
and the relations between them.

\section{Observations}
\label{s.obs}

We analyse 35 optical spectra of Be stars obtained with 
the ESpeRo Echelle spectrograph \citep{2017BlgAJ..26...67B}   
on the 2.0~m telescope of the Rozhen National Astronomical Observatory, Bulgaria, 
and 27 spectra from the ELODIE archive \citep{2008asvo.proc...47I},
obtained with the 1.93~m telescope of Observatoire de Haute-Provence, France. 
Both instruments are fiber-fed Echelle spectrographs 
providing high resolution spectra in the optical range. 
The ESpeRo spectrograph has a resolution of $\sim$ 30~000 and covers the range 3900~\AA\
-- 9000~\AA. 
The ELODIE spectrograph has
a resolution $\sim$ 42~000 and covers the range 3895~\AA\ -- 6815~\AA. 
The spectra Rozhen spectra are obtained during the period 2015 -- 2023, the ELODIE -- 
1997-2003. 
Some examples of the profiles of H$\alpha$, H$\beta$ and H$\gamma$ emission lines are shown on Fig.~\ref{f.ex}.

On the spectra, we measure the distance between the peaks of the Balmer emission lines
H$\alpha$ ($\Delta V_\alpha$),
H$\beta$  ($\Delta V_\beta$),
H$\gamma$  ($\Delta V_\gamma$). 
The measurements are listed in Table~\ref{t.Roz} and Table~\ref{t.Elo}.
These parameters are identical to $(\Delta \lambda)_s$, as shown on Fig.~3 of 
\cite{1976IAUS...70..123S}. 
To measure the position, we applied Gaussian fitting at the top of each peak,
as shown in  Fig.~\ref{f.6} in the Appendix. 
The typical errors are $\pm 5$~km~s$^{-1}$ for $\Delta V_\alpha$, 
$\pm 7$~km~s$^{-1}$ for $\Delta V_\beta$, 
and  $\pm 10$~km~s$^{-1}$ for $\Delta V_\gamma$.

\section{Relations}
\label{s.rel}

For the Be stars, \citet{1988A&A...189..147H} 
find that the peak separations of H$\beta$ and H$\alpha$ emission lines
follow  approximately the relation  $\Delta V_\beta \approx 1.8 \Delta V_\alpha$.
\cite{1992A&AS...95..437D} find that average  
$\Delta V_\beta / \Delta V_\alpha = 1.6 - 1.8$.
Using more data, we have found 
that this relation is not valid above $\Delta V_\alpha \approx 200$~km~s$^{-1}$, and a 
linear fit of the type $y=a+bx$ is more appropriate. 
In Fig.~\ref{f.ab} we plot $\Delta V_\beta$ versus $\Delta V_\alpha$ for 56 Be stars.
In this figure the black circles are form \citet{1988A&A...189..147H},
\cite{1992A&AS...95..437D}, and \citet{2013A&A...550A..79C}. 
The green squares are ELODIE spectra (Table~\ref{t.Elo}). 
The red pluses are Rozhen data. The Rozhen data include 
the new spectra from Table~\ref{t.Roz}, and the published spectra 
obtained with the same telescope setup \citep{2016A&A...593A..97Z, 2022AN....34324019Z}. 
We find the following relationship between $\Delta V_\alpha$ and $\Delta V_\beta$:
\begin{equation}
\Delta V_\beta= 1.01(\pm 0.01) \; \Delta V_\alpha + 64.2(\pm2.5)\; \; {\rm km \; s^{-1}}
\label{eq.ab}
\end{equation}
This relation is valid for the range of $\Delta V_\alpha$ from 40~km~s$^{-1}$
to 530~km~s$^{-1}$.
It is very similar to our finding for a smaller sample 
and smaller  range \citep{2022AN....34324019Z}.  
As expected,  
there is a very strong correlation between these two quantities, 
with Pearson correlation coefficient 0.95,
Spearman's rank correlation 0.93, 
and significance $10^{-20}$.

In Fig.~\ref{f.bg} we plot $\Delta V_\gamma$ versus $\Delta V_\beta$. 
The black circles are data 
from \citet{1976IAUS...70..123S} and \citet{1988A&A...189..147H}. 
The green  squares are ELODIE data, the red pluses - Rozhen data. 
For the Be stars, \citet{1988A&A...189..147H} 
find that the peak separations of H$\beta$ and H$\gamma$ emission lines
are connected as $\Delta V_\gamma = 1.2 \Delta V_\beta$. 
Here using more data, we find the following connection  between
$\Delta V_\beta$  and $\Delta V_\gamma$:
\begin{equation}
\Delta V_\gamma =  1.034 (\pm 0.031) \;  \Delta V_\beta + 14.9(\pm 4.2) \; \; {\rm km \; s^{-1}}
\label{eq.bg}
\end{equation}
This relation is valid for the range of $\Delta V_\beta$ from 20~km~s$^{-1}$
to 340~km~s$^{-1}$.  There is a very strong correlation between these two quantities, 
with Pearson correlation coefficient 0.97, Spearman's rank correlation 0.96, 
and significance $10^{-18}$.

For 14 objects we have $\Delta V_\gamma$ and $\Delta V_\alpha$ 
on the same spectrum, in total 36 data points,  
because some objects are observed several times. 
In Fig.~\ref{f.ag} we plot $\Delta V_\gamma$ versus  $\Delta V_\alpha$. 
These data points refer to spectra on which both  $\Delta V_\alpha$ and $\Delta V_\gamma$
can be measured.  
We are not able to measure $\Delta V_\alpha$ and $\Delta V_\gamma$ on each spectrum because 
(1) if the $H\alpha$ emission is strong, the H$\alpha$ peaks are very close one to another and can merge; 
(2) if the $H\alpha$ emission is weak, the emission peaks in H$\gamma$ are not visible; 
(3) the sensitivity of the ESpeRo spectrograph is lower at  H$\gamma$. 
From  Eq.~\ref{eq.ab} and Eq.\ref{eq.bg}
we expect the following connection between 
$\Delta V_\gamma$ and  $\Delta V_\alpha$:  
\begin{equation}
  \Delta V_\gamma = 1.04 \; \Delta V_\alpha + 81.8 \; \; {\rm km \; s^{-1}}
\label{eq.ag1}
\end{equation}
This equation is represented as a  blue dashed line in Fig.~\ref{f.ag}. 

Because there are some objects with one measurement and a few objects with 3-4 measurements, 
we performed the linear fit, using different subsamples (bootstrap technique, e.g. \cite{1979.Efron})
in a way to have in the subsamples 1 or 2 measurements for each object
and thus not to give too much weight to  the objects with 4 measurements.
A linear fit in the form $y=ax+b$ gives  
coefficients in the range $0.98 < a < 1.11$ and $71.3 \le b \le 82.6$, 
with average values  $a=1.065$ and  $b=74.8$: 
\begin{equation}
  \Delta V_\gamma = 1.065 (\pm 0.062) \; \Delta V_\alpha + 74.8 (\pm 6.2) \; \; {\rm km \; s^{-1}}
\label{eq.ag2}
\end{equation}
This fit is represented as a black solid line in Fig.~\ref{f.ag}. 
The correlation is strong with Pearson correlation coefficient 0.90, 
Spearman's rank correlation 0.91, and significance $10^{-9}$.

The black solid line (derived from the fit over the data) and 
the blue dashed line (based on the relationships  $\Delta V_\alpha$ vs. $\Delta V_\beta$ and  
$\Delta V_\beta$ vs. $\Delta V_\gamma$) are similar, 
which confirms that the results are selfconsistent. 
In all three cases ($\Delta V_\beta$ versus $\Delta V_\alpha$, 
$\Delta V_\gamma$ versus $\Delta V_\beta$, and
$\Delta V_\gamma$ versus $\Delta V_\alpha$,  
presented in Fig.~\ref{f.ab}, \ref{f.bg}, and \ref{f.ag}, respectively)  
we find a linear relationship of the type  $y= ax + b$, with $a \approx 1.0$.

\section{Disc size}
\label{Rd1}

In the Be stars, the distance between the peaks of  H$\alpha$ emission line 
can be regarded as a measure of 
the outer radius ($R_{disc}$) of the emitting disc \citep[e.g.][]{1972ApJ...171..549H}. 
 \begin{equation}
   R_{disc} = R_1 \left( \frac{2 \; v \sin{i}}{(1-\epsilon) \; \Delta V\alpha}\right)^2, 
  \label{Huang3}
  \end{equation}
where $G$ is the gravitational constant,  
$M_1$ is the mass of the Be star, $v\,\sin{i}$ is the projected rotational velocity,
$\sin{i}$ is the inclination to the line of sight. 
The term $(1-\epsilon)$ represents the fact that 
the Be stars are rotating below the critical value \citep{2003PASP..115.1153P}. 

The relations between $\Delta V_\gamma$, $\Delta V_\beta$, and $\Delta V_\alpha$ 
give us the possibility to estimate $ R_{disc}$, when two peaks 
are visible in $H\beta$, 

 \begin{equation}
   R_{disc} = R_1 \left( \frac{1.01 * 2 \; v \sin{i}}
                              {(1-\epsilon) \; (\Delta V_\beta - 64.2) }  
	          \right)^2, 
  \label{eq.Rab}
  \end{equation}
or in $H\gamma$ emission:
 \begin{equation}
   R_{disc} = R_1 \left( \frac{1.065 * 2 \; v \sin{i}}
                              {(1-\epsilon) \; (\Delta V_\gamma - 74.8) }  
	          \right)^2,  
  \label{eq.Rag}
  \end{equation}
where $\Delta V_\beta$,  $\Delta V_\gamma$, and $v \sin{i}$
are in km~s$^{-1}$. 

Radius estimation through the method 
of \cite{1972ApJ...171..549H} is a good approximation for symmetric profiles
with double peaked $H\alpha$. If the two peaks of $H\alpha$ are not clearly visible, 
the equivalent width of H$\alpha$ emission can be used instead
[e.g. Section~3 of \cite{2006MNRAS.368..447C}, and Eq.~5 in \cite{2022AN....34324019Z}].
The relationships given above by Eq.~\ref{eq.Rab} and Eq.~\ref{eq.Rag} 
offer a third possibility.

\section{Discussion}

Recently, \cite{2018ApJ...853..156W} and \cite{2021AJ....161..248W},
using the UV spectra from HST and IUE,  searched and found hot subdwarf companions of Be stars.   
Their results indicate that probably the rapid rotation of most Be stars is a result of 
mass transfer in a close binary system, as suggested by
\citet{1991A&A...241..419P}. 
In this scenario, many Be stars are expected to have companions
that are the remnants of the mass donor. 
The donor star might be stripped and
become a hot subdwarf star in a Be+sdO binary, 
it  might explode to create a neutron star or black hole 
in a Be/X-ray binary, 
or the binary might be disrupted by the supernova 
explosion \citep{2014LRR....17....3P}.    

Potentially, the deviations of individual objects 
from the average behaviour 
(as expressed in Eq.~\ref{eq.ab}, Eq.~\ref{eq.bg}, and  Eq.~\ref{eq.ag2}) 
can be used to investigate 
the influence of the secondary on the Be disc structure, 
disc truncation, 
(because the truncation occurs only at certain radii;  
\cite{2001A&A...377..161O}), 
warping of the disc  \citep{2011MNRAS.416.2827M}, etc. 
The correlations can also  be useful to test the theoretical models 
for formation of the emission lines 
in Be discs like those presented in Fig.~14 of \cite{2008PASJ...60..749I}. 

The Be stars show spectral variability both on short time-scales -- hours to months 
\citep{2003PASP..115.1153P, 2017JApA...38....6P}  
and long time-scales --  up to years and decades 
\citep{1994A&AS..108..237M}.   
One major observational aspect of Be stars is that some of them 
change phase from Be $-$ Be-shell $-$ normal B star $-$ Be. 
It will be interesting to follow the evolution of  Be stars on these diagrams during
the phase changes.

\section{Conclusions}
We analysed high resolution spectra of Be stars. 
We find the  relationships  between separations of the peaks
of the H$\alpha$, H$\beta$, H$\gamma$ emission lines. 
The correlations found should be 
useful for future theoretical modeling and better understanding 
of the  circumstellar discs of the Be stars in general. 

\vskip 0.5cm 

{\bf Acknowledgments: }
This work was supported by the  \fundingAgency{Bulgarian National Science Fund}   
\fundingNumber{project KP-06-H28/2 08.12.2018  "Binary stars with compact object"}.

\vskip 0.5cm 
{\bf Conflict of interest: }
The authors declare no potential conflict of interests.

\clearpage

\bibliography{zaman.ref4.bib}






\begin{table}
\small
\centering
\caption{Rozhen spectra -- $\Delta V_\alpha$, $\Delta V_\beta$ and $\Delta V_\gamma$.  } 
\begin{tabular}{cclrrrcc}
\hline 
object          &     &    file   & $\Delta V_\alpha$ & $\Delta V_\beta$ & $\Delta V_\gamma$ & \\
                &     &  	 & [km s$^{-1}$]     & [km s$^{-1}$]	& [km s$^{-1}$]     & \\ 
BD$-$00 3543 & HD 173371   & 20220520.01.fit	 &    223.4 &	  278.7   &  ---    &   &\\  
BD$-$00 3543 & HD 173371   & 20220520.02.fit	 &    223.7 &	  281.4   &  ---    &   &\\  
BD$-$00 3543 & HD 173371   & 20220512.1800s.fit	 &    218.2 &	  269.3   &  ---    &   &\\  
BD$+$02 3815 & HD 179343   & 20220512.1800s..fit   &    227.4 &	  270.7   &  ---    &   &\\  
BD$+$02 3815 & HD 179343   & 20220521.01.fit	 &    230.2 &	  277.1   &  ---    &   &\\    
BD$+$02 3815 & HD 179343   & 20220521.02.fit	 &    229.9 &	  278.3   &	    &   &\\  
BD$+$05 3704 & HD 168797   & 20220520.01.fit	 &    549.6 &	  651.1   &	    &   &\\  
BD$+$05 3704 & HD 168797   & 20220520.02.fit	 &    521.1 &	  658.3   &	    &   &\\  
BD+30 0591   & X Per       & 20151223.1.fit        &    102.9 &	  143.2   & 168.4   &  &\\
BD+30 0591   & X Per       & 20160130.fit          &     96.8 &	  127.0   & 135.7   &  &\\ 
BD+30 0591   & X Per       & 20161211.fit          &     59.2 &     132.7   & 154.3   &  &\\ 
BD+30 0591   & X Per       & 20171230.fit          &    127.0 &	  150.6   & 163.5   &  &\\ 
BD$+$37 0675 & HD 18552    & 20220120.1.120s.fit   &    151.2 &	  234.2   &	    &   &\\  
BD$+$37 0675 & HD 18552    & 20220120.1.600s.fit   &    151.6 &	  228.3   &   257.9 &   &\\  
BD$+$37 0675 & HD 18552    & 20220120.2.120s.fit   &    150.2 &	  236.5   &	    &   &\\ 
BD$+$37 0675 & HD 18552    & 20220120.2.600s.fit   &    150.1 &	  230.9   &   280.7 &   &\\  
BD$+$47 0857 & psi Per     & 20220120.1.120s.fit   &    133.4 &	  ---	  &	    &   &\\  
BD$+$47 0857 & psi Per     & 20220120.1.300s.fit   &    132.2 &	  174.0   &   205.1 &   &\\  
BD$+$47 0857 & psi Per     & 20220120.2.120s.fit   &    132.9 &	  174.8   &   201.2 &   &\\  
BD$+$47 0857 & psi Per     & 20220120.2.300s.fit   &    131.7 &	  176.5   &   206.9 &   &\\  
BD$+$50 0825 & HD 23552    & 20220320.1.20min.fit  &    184.0 &	  261.2   &	    &   &\\  
BD$+$50 0825 & HD 23552    & 20220320.2.20min.fit  &    187.0 &	  245.1   &	    &   &\\  
BD$+$40 1213 & HD 33604    & 20220119.1.20min.fit  &    ---   &	   29.8   &   39.7  &   &\\  
BD$+$40 1213 & HD 33604    & 20220119.2.20min.fit  &    ---   &	   26.5   &   48.7  &   &\\  
BD$+$42 1376 & HD 37657    & 20220119.1.20min.fit  &    412.7 &	  467.4   &  ---    &   &\\  
BD$+$42 1376 & HD 37657    & 20220119.2.20min.fit  &    404.2 &	  421.1   &  ---    &   &\\  
CSI+44-04374 & LS V +44 17 & 20161211.fit          &    290.6 &     346.9   &  ---    &  &\\
CSI+44-04374 & LS V +44 17 & 60m.1.20230211.fit    &    250.8 &     338.9   &  ---    &  &\\ 
BD+53 2790   & 4U 2206+54  & 20160620.fit        &    467.0 &     503.4   &  ---    &  &\\ 
BD+53 2790   & 4U 2206+54  & 20151227.fit        &    493.1 &     582.7   &  ---    &  &\\
BD+53 2790   & 4U 2206+54  & 20190819.fit        &    516.4 &     599.7   &  ---    &  &\\
BD+53 2790   & 4U 2206+54  & 20200905.fit        &    493.0 &     525.4   &  ---    &  &\\
CSI+59-01302 & LS I +59 79 & 20160621.fit        &    177.4 &     265.1   &  ---    &   & \\
CSI+59-01302 & LS I +59 79 & 20160923.fit        &    179.5 &     249.5   &  ---    &   & \\
BD+59 0144   & gamma Cas   & 20160621.1.fit      &    ---   &     138.7   &  176.2  &   & \\  
BD+59 0144   & gamma Cas   & 20160621.2.fit      &    ---   &     132.0   &  172.3  &   & \\
\hline								
 \end{tabular} 						  
 \label{t.Roz} 
$\; $  \\
$\; $  \\
$\; $  \\				    
\centering
\caption{ELODIE spectra -- $\Delta V_\alpha$, $\Delta V_\beta$ and $\Delta V_\gamma$.  } 
\begin{tabular}{cccrrrcc}
\hline 
  object     &              & file name                  &   $\Delta V_\alpha$ & $\Delta V_\beta$ & $\Delta V_\gamma$ & \\
             &              &                            &   [km s$^{-1}$] & [km s$^{-1}$] & [km s$^{-1}$] &  \\ 
		       
 BD$+$00 4872  &  pi Aqr      & elodie\_20011222\_0005.fits  &  563.8  &  ---   & ---   & \\ 
 BD$+$04 1002  &  omega Ori   & elodie\_19981123\_0035.fits  &  244.1  &  339.8 & 370.1 & \\
 BD$+$04 1002  &  omega Ori   & elodie\_19981125\_0014.fits  &  242.1  &  346.0 & 366.9 & \\
 BD$+$04 1002  &  omega Ori   & elodie\_19981126\_0032.fits  &  242.4  &  350.6 & 348.0 & \\
 BD$+$04 1002  &  omega Ori   & elodie\_19981125\_0033.fits  &  245.2  &  333.1 & 340.3 & \\
 BD$+$04 3570  &  66 Oph      & elodie\_20030808\_0013.fits  &  142.5  &  211.8 & 243.9 & \\	   
 BD$+$04 3570  &  66 Oph      & elodie\_20030811\_0021.fits  &  140.0  &  214.3 & 253.2 & \\
 BD$+$04 3570  &  66 Oph      & elodie\_20030819\_0014.fits  &  142.2  &  218.4 & 234.4 & \\
 BD$+$08 1774  &  beta CMi    & elodie\_20020328\_0018.fits  &  133.3  &  215.3 & ---	& \\
 BD$+$11 4784  &  31 Peg      & elodie\_20011222\_0006.fits  &   36.7  &   67.9 &  92.3 & \\
 BD$+$23 0558  &  28 Tau      & elodie\_20011121\_0042.fits  &   91.0  &  150.2 & 181.4 & \\
 BD$+$23 0558  &  28 Tau      & elodie\_20011220\_0013.fits  &   98.8  &  139.0 & 167.2 & \\
 BD$+$23 0558  &  28 Tau      & elodie\_20030124\_0005.fits  &  105.7  &  153.6 & 189.8 & \\
 BD+30 0591    &  X Per       & elodie\_20041117\_0010.fits  &	110.8  &  167.4	& 189.5 & \\
 BD$+$34 4371  &  ups Cyg     & elodie\_20030809\_0026.fits  &   59.1  &  106.8 & 153.7 & \\
 BD$+$34 4371  &  ups Cyg     & elodie\_20030811\_0030.fits  &   45.1  &  107.8 & 148.6 & \\
 BD$+$34 4371  &  ups Cyg     & elodie\_20030819\_0022.fits  &   56.0  &  110.8 & 148.2 & \\
 BD$+$47 0857  &  psi Per     & elodie\_20030814\_0023.fits  &  126.8  &  173.1 & 187.7 & \\
 BD$+$47 0857  &  psi Per     & elodie\_20030815\_0025.fits  &  134.5  &  172.7 & 193.8 & \\
 BD$+$47 0857  &  psi Per     & elodie\_20030819\_0046.fits  &  134.9  &  172.9 & 192.4 & \\
 BD$+$49 0444  &  phi Per     & elodie\_19970917\_0023.fits  &  194.5  &  218.7 & 234.9 & \\ 
 BD$+$49 0444  &  phi Per     & elodie\_19970917\_0024.fits  &  189.8  &  220.8 & 242.2 & \\
 BD$+$64 1527  &  6 Cep       & elodie\_20030818\_0019.fits  &	65.4   &  136.8 & 177.3 & \\
 BD$+$64 1527  &  6 Cep       & elodie\_20030819\_0017.fits  &	65.6   &  132.3 & 172.7 & \\
 BD$+$64 1527  &  6 Cep       & elodie\_20030819\_0025.fits  &	66.0   &  133.2 & 174.0 & \\    
 BD$+$70 0703  &  kappa Dra   & elodie\_20000126\_0038.fits  &  116.5  &  198.1 & 212.2 & \\
 BD$+$70 0703  &  kappa Dra   & elodie\_20000128\_0020.fits  &  100.7  &  200.0 & 216.3 & \\
 \\			    	   
 \hline 
 \end{tabular}       		        			     
 \label{t.Elo}    		        	   
 \end{table}	     				    

\appendix

\section{Supporting information\label{app1}}

  \begin{figure}[ht!]  
   \vspace{8.9cm} 
  \includegraphics{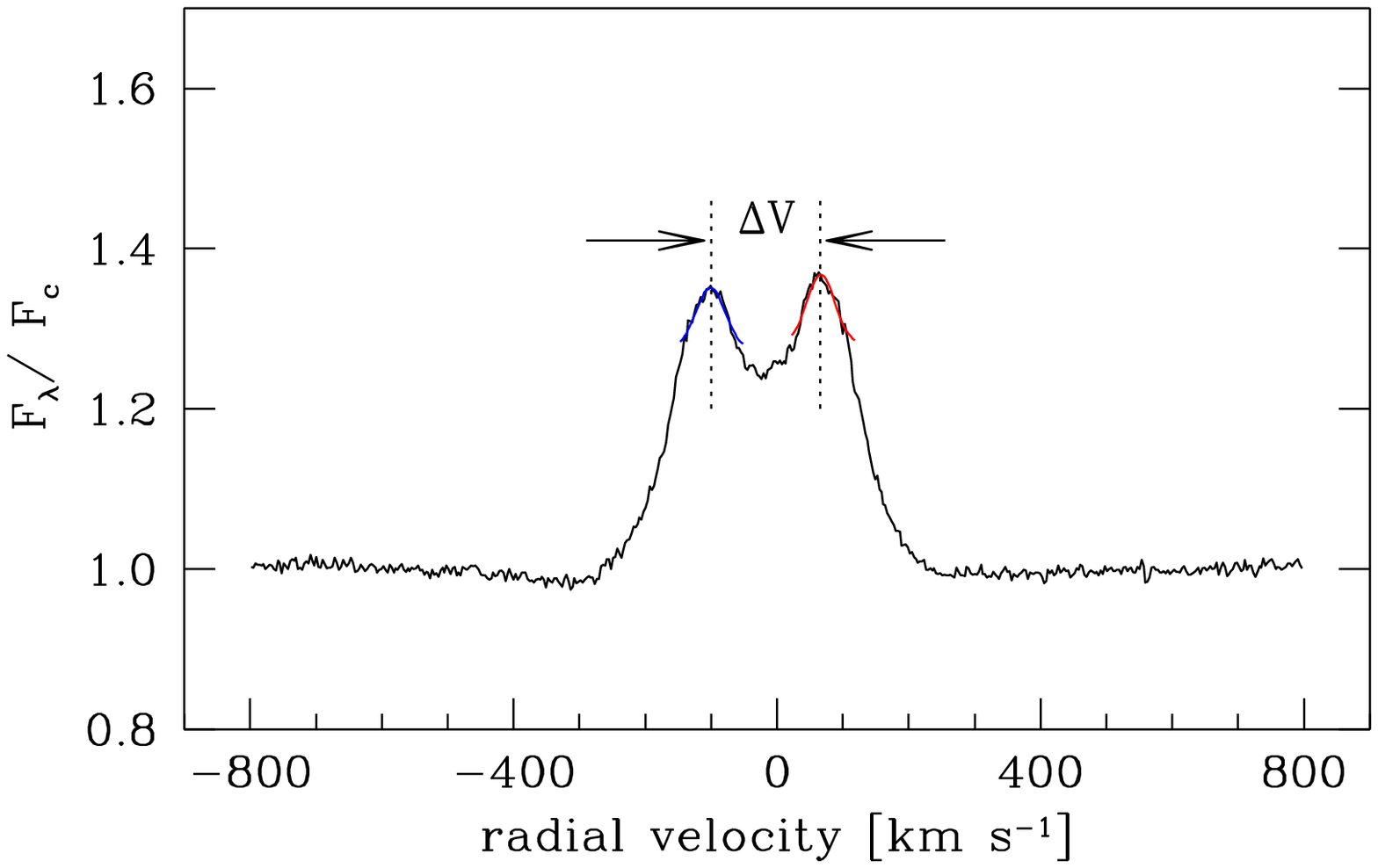} 	  
  \caption[]{Gaussian fits in the upper part of the emission line are used to find the position
             of the violet and red peaks.  
	     The distance between the peaks, $\Delta V$, is also marked. } 
  \label{f.6}  
\end{figure}

%
%

\end{document}